\documentclass[reviewcopy]{elsart}

\usepackage{amsmath}
\usepackage{amsfonts}
\usepackage{amssymb}
\usepackage [ansinew] {inputenc}
\usepackage[dvips]{graphicx}
\usepackage[nomarkers,nolists]{endfloat}
\usepackage[square, comma, sort&compress]{natbib}

\begin{document}

\begin{frontmatter}
\title{Coupling theory for counterion distributions based in Tsallis statistics}
\author{V.~Garc\'{\i}a-Morales\corauthref{cor1}}
\ead{vladimir.garcia@uv.es}, \author{J.~Cervera}, \author{J.~Pellicer}
\corauth[cor1]{Corresponding author. Tel: +34 96 354 3119; Fax: +34 96 354 3385}
\address{Departament de Termodin\`{a}mica, Universitat de
Val\`{e}ncia, C/Dr. Moliner 50, E-46100 Burjassot, Spain}
\begin{abstract}
\noindent{It is well known that the Poisson-Boltzmann (PB) equation yields the
exact counterion density around charged objects in the weak coupling limit.
In this paper we generalize the PB approach to account for coupling
of arbitrary strength by making use of Tsallis q-exponential distributions. 
Both the weak coupling and the strong coupling limits are reproduced. 
For arbitrary coupling we also provide 
simple analytical expressions which are compared to recent Monte Carlo simulations 
by A. G. Moreira and R. R. Netz [Europhys. Lett. 52 (2000) 705]. 
Excellent agreement with these is obtained.}
\end{abstract}
\end{frontmatter}

\section{Introduction}

Many biopolymers, as DNA and actin, are strongly charged
and the explanation of the screening of their charges by counterions in solution
is of large relevance (for a recent review, see \cite{Grosberg}).
A detailed knowledge of these screening effects not only allows to determine properties of individual
polyelectrolytes, but also forces acting between them and the thermodynamic and transport properties
of their solutions \cite{Grosberg, Perel, Shk1, Moreira, Nguyen}. The traditional Poisson-Boltzmann 
(PB) equation obtained from Boltzmann-Gibbs (BG) thermostatistics has long been known 
to fail to explain the concentration profiles of multivalent counterions 
close to a charged macroion \cite{Bo, Perel}. The Perel-Shklovskii (PS) theory \cite{Perel}
based in the strongly correlated liquid and the Wigner crystal \cite{Shk2} concepts 
has provided a qualitative explanation of this failure
and also some quantitative results that have been recently confirmed by numerical simulations 
\cite{Messina, Tanaka}.  
The approach has been succesfully applied
to explain charge inversion. However, an analytical expression for the counterion concentration profile
valid for the whole spatial range and for any ionic number $Z$ has not yet been proposed.
The recent theory of Moreira and Netz \cite{Moreira2} making use of field-theory methods, 
rigorously establishes the strong coupling limit for the counterion distribution besides a charged planar wall. 
The expression provided in this limit by the authors agrees with their Monte Carlo simulations, 
which are also carried for moderate coupling \cite{Moreira2}. For the weak coupling regime there exists also density 
functional approaches \cite{Stevens} and methods involving integral equations \cite{Marcelja} 
that provide accurate results. However, a systematic and unified 
theory is still lacking for coupling of arbitrary strength.

In the present paper we attempt to provide such a theory. The theory 
exactly reproduces both the weak and strong coupling limits being also consistent with classical rigorous results.
One of the main advantages of the theory is that strikingly simple expressions are provided for arbitrary coupling 
in excellent agreement with Monte Carlo simulations by Moreira and Netz \cite{Moreira2}.
The PB equation is generalized to account for the correlations making use of the Tsallis q-exponential distributions. 

We present first results of the classical PB theory. Let us consider a massive insulating macroion
having a negative surface charge density $-\sigma$. Assume further that the only counterions in solution
are those dissociated from the surface so that the system macroion+solution is neutral. When there are
no charge correlations (monovalent counterions and low $\sigma$) the mean field PB equation
\begin{equation}
\Delta \psi=-4\pi l N(0) e^{-\psi} \label{PBE}
\end{equation}
describes accurately the counterion concentration profile that screens the charge of the macroion. In Eq.(\ref{PBE})
$\psi=-\beta eZ[\phi(x)-\phi(0)]$ ($\beta=1/kT$ is the room Lagrange parameter, $e$ is the electron charge
and $Z$ is the ionic number) $N(0)$ and $\phi(0)$ being the counterion concentration
and the electric potential at the surface of a charged object respectively.
In Eq. (\ref{PBE}), $l \equiv Z^{2}l_{B}$, where 
$l_{B}=e^{2}/DkT$ is the Bjerrum length, which is a characteristic threshold for repulsion forces.
The relative permittivity for water is $D \approx 80$. Eq.(\ref{PBE}) results from introducing
in the Poisson equation $\Delta \psi=-4\pi l N(x)$ the
Boltzmann equation
\begin{equation}
N(x)=N(0)e^{-\psi} \label{Boltzy}
\end{equation}
which relates the counterion concentration at a position $x$ to that in the vicinity
of the surface, considering a planar geometry. Two boundary conditions are to be supplemented
to Eq.(\ref{PBE})
\begin{eqnarray}
\left.\frac{d\psi}{dx}\right|_{x=0}&=&\frac{2}{\lambda} \label{BC1} \\
\left.\frac{d\psi}{dx}\right|_{x \to \infty}&=& 0 \label{BC2}
\end{eqnarray}
where $\lambda=D/(2\pi \beta \sigma Z e)$ is the Gouy-Chapman length \cite{Grosberg}. 
Eqs.(\ref{PBE}), (\ref{BC1}) and
(\ref{BC2}) completely specify the boundary value problem. Its solution is 
\begin{eqnarray}
N(x)=\frac{1}{2\pi l} \frac{1}{(x+\lambda)^2} \label{PBEprofile}
\end{eqnarray}
This profile correctly describes the counterion density in the weak coupling limit. Out of this limit, significant departures (of importance for many phenomena) occur. It has been shown recently that
in the limit of strong coupling, the counterion concentration profile decays exponentially with the
distance to the macroion surface
\begin{equation}
N(x)=\frac{1}{2\pi l\lambda^{2}} \exp\left(-\frac{x}{\lambda}\right) \label{More}
\end{equation}
The mean field PB approach cannot account for this profile. Furthermore, although extensive
Monte Carlo calculations have been performed for this system \cite{Moreira2}, a theory for
arbitrary coupling is still lacking.

\section{Generalized mean field theory}

The exact evaluation of the 
Gibbs canonical partition function for the many (interacting) particle system 
provides the best description of the counterion distributions. However, an analytical closed form for 
this canonical partition function is inviable and requires computational resources in most cases.
The PB theory above outlined is a valid approximation when correlations are not important and 
provides quite useful expressions strictly valid only in the weak coupling limit.

We can wonder if a mean field description can still be possible when coupling of arbitrary strength is incorporated. 
If one assumes a mean field potential to hold, its value must then be allowed to fluctuate. 
We can assume that fluctuations of the mean field value
of the electrostatic potential are distributed according to a $\chi^{2}$ (or gamma) 
probability distribution function of order $n$, $f(\psi')$  
\begin{equation}
f(\psi')=\frac{1}{\Gamma\left(\frac{n}{2}\right)}\left\{\frac{n}{2\psi}\right\}^{\frac{n}{2}}\psi'^{\frac{n}{2}-1}
\exp\left\{-\frac{n\psi'}{2\psi}\right\}
\end{equation}
Then, all possible values for the mean field value of 
the electrostatic potential must be weighted by this distribution and the canonical partition function
becomes \cite{Wilk}
\begin{equation}
\int_{0}^{\infty}f(\psi')e^{-\frac{\widetilde{\beta}}{\beta}\psi'}d\psi'=\left(1-(1-q)\widetilde{\beta}\psi/\beta \right)^{\frac{1}{1-q}}
\end{equation}
which coincides with the Tsallis q-exponential distribution (here $q\equiv 1+2/n$). 
These expressions hold rigorously for $q > 1$ although for $q < 1$ there exists 
an analogous treatment \cite{Wilk}. It is to be noted, however, than in previous considerations
the fluctuating variable is the Lagrange parameter and \emph{not} the energy. 
We consider here fluctuations in energy in a rather formal way due to the fact that energy
is still approximated to a mean field value beyond its traditional domain. Fluctuations
of arbitrary strenght must hence be included. $\widetilde{\beta}$ 
is proportional to the room Lagrange parameter $\beta$ 
(i.e. the Boltzmann-Gibbs Lagrange parameter) and related through a constant $k_{q}$ which can
depend on $q$. Clearly, if $q=1$ in this expression then $f(\psi')=\delta(\psi'-\psi)$ 
and the Boltzmann factor corresponding to the PB mean field theory is regained. 
$q$ is the Tsallis entropic parameter, which is related in the present problem
to the coupling constant entering in the Hamiltonian. 

For coupling of arbitrary strength Eq.(\ref{Boltzy}) can then be generalized in the following way
\begin{equation}
N(x)=N(0)e_{q}^{-\widetilde{\beta}\psi/\beta}=
N(0) \left[1-(1-q)\widetilde{\beta}\psi/\beta \right]^{\frac{1}{1-q}} \label{Tsally}
\end{equation}
In this work $q$ is restricted to the range $0 \le q \le 1$. 
When $q=1$, the Boltzmann equation, Eq. (\ref{Boltzy}) is regained.

q-exponential distributions are extremely interesting and have proven many applications \cite{web}. 
These can be obtained directly by maximizing the Tsallis entropic form $S_{q}$ \cite{Tsal1}
\begin{equation}
S_{q}=k\frac{\sum_{j}^{\Omega}p_{j}^{q}-1}{1-q}
\end{equation}
imposing the following constraint for the biased average of the energy \cite{Tsal2}
\begin{equation}
\frac{\sum_{j}^{\Omega}p_{j}^{q}\epsilon_{j}}{\sum_{j}^{\Omega}p_{j}^{q}}=U
\end{equation}
The q-exponential distributions have been applied succesfully 
to systems in which some degree of fractality is present, 
and are used here tentatively provided that correlated structures possess fractal features as it is well known, 
for example, from studies of critical phase transitions. 

It is important to note that it is not the validity of BG thermostatistics what is at issue here. As indicated above, 
BG thermostatistics provides the best description of the physics involved. 
What we suggest is to generalize the PB mean field theory, which is based on the factorization of the many particle
probability distributions. When correlations are important, this factorization cannot be carried out in the framework
of BG thermostatistics. It is proposed that the factorization in terms of single particle probability distributions 
is, indeed, still possible if one uses the Tsallis generalized framework, considering the appropriate probability distributions. Since a 
mean value for the electrostatic potential is still assumed to hold, it must be allowed to strongly fluctuate
and this is what we have done above by assuming a chi-square probability distribution for the fluctuations.

Tsallis statistics has proven a valuable tool in dealing with long range interactions. 
The existence of screening in the case of Coulomb systems 
and, hence, of a thermodynamic limit \cite{Lieb} is not, in principle, 
a matter for worrying about concerning its application. 
It is known, for example, that the Tsallis canonical 
partition function can be derived assuming that a certain 
hierarchy of structures (for example a polydisperse distribution of particle sizes) 
is present with a given probability \cite{Johal0}. The interaction between these structures, 
considered at the ensemble level, has a thermodynamic limit. 
In our system, the structures could be understood as those composed by each counterion 
and a number of shells of its screening atmosphere. The mean field theory here is not applied 
directly to each counterion moving in a potential created by the others, 
but to a structure containing the counterion and a portion of screening 
atmosphere of averaged size (which is mediated by 
the surface charge and the ionic valence and is contained in the q-dependence). 
In this framework $q \approx 1$ accounts for disordered screening shells corresponding to low charged ions 
(and low surface charge concentration) whereas $q \ll 1$ describes highly charged 
charged systems with screening shells having some degree of order. 
These screening shells are of fractal character. 
Order enters in the present description by remembering that the parameter $q$ tends
to privilege certain microstates again others between all available ones. Interestingly, a strongly
correlated liquid has been suggested to form a Wigner crystal at the macroion surface \cite{Shk1}, 
implying favoring certain concrete microstates against all those allowed in the disordered bulk phase.

In the theory developed here the Tsallis canonical partition function is hence to 
be considered as the correct way of factorizing 
the many-particle distribution of the exact BG canonical partition function in single-particle 
(mean field) contributions when the pair interaction term is considered. 
The q-dependence and the structure of the Tsallis canonical partition function corresponds 
then to a more general averaging which includes correlations in a natural way. 

It is also worth noting that there exists also an extensive approach \cite{Johal} that leads 
to Tsallis q-exponential distributions. This is of interest to establish a variational principle following 
analogous lines to those of Ref. \cite{Deserno}. It leads to the 
following generalized PB equation
\begin{equation}
\Delta \psi=-4\pi \widetilde{l} N(0) e_{q}^{-\widetilde{\beta}\psi/\beta}=
-4\pi l N(0) \left[1-(1-q)\widetilde{\beta}\psi/\beta \right]^{\frac{1}{1-q}} \label{GPBE}
\end{equation}
which can also be easily obtained by replacing Eq. (\ref{Tsally}) in the Poisson equation.

The boundary value problem specified by Eqs.(\ref{BC1}), (\ref{BC2}) and (\ref{GPBE})
can now be analytically solved in an analogous way to that followed
in the PB approach, which led to Eq. (\ref{PBEprofile}). The counterion concentration profile
is then given by 
\begin{equation}
N(x)=\frac{(2-q)\widetilde{\beta}}{2\pi l \beta \lambda^{2}}\left(1+\frac{q\widetilde{\beta}x}{\beta \lambda}\right)^{-2/q} \label{profil}
\end{equation}
As can be observed, for $x=0$, the counterion concentration is
\begin{equation}
N(0)=\frac{2-q}{2\pi l\lambda^{2}}\frac{\widetilde{\beta}}{\beta} \label{N0}
\end{equation}
An exact result obtained with the classical theory and which must also hold here comes
from the contact theorem \cite{Wenner} which states that $N(0)=1/(2\pi l \lambda^{2})$.
This allows to know $\widetilde{\beta}$ in terms of the actual Lagrange parameter $\beta$
thus removing the arbitrariness of this parameter. We obtain
\begin{equation}
\widetilde{\beta}=\frac{\beta}{2-q} \label{beta}
\end{equation}
and, therefore, the counterion concentration profile can be finally written as
\begin{equation}
N(x)=\frac{1}{2\pi l \lambda^{2}}\left(1+\frac{qx}{(2-q) \lambda}\right)^{-2/q} \label{profile}
\end{equation}
As can be seen, for $q=1$ (weak-coupling limit) Eq.(\ref{PBEprofile}) is regained. 
The counterion concentration profile is normalized \emph{independently of q}
\begin{equation}
\int_{0}^{\infty}N(x)dx=\frac{1}{2\pi l\lambda}=\frac{\sigma}{Ze} \label{normalization}
\end{equation}
In the strong coupling case $(q \to 0)$ Eq. (\ref{profile}) describes a 
pure exponential decay
\begin{equation}
N(x)=\frac{1}{2\pi l\lambda^{2}} \exp\left(-\frac{x}{\lambda}\right) \label{profile0}
\end{equation}
in agreement with the strong-coupling theory by Moreira and Netz, Eq. (\ref{More}).

Since the whole system is neutral, the apparent surface charge density 
at position $x$ can be calculated by making use of Eq.(\ref{profile}) as
\begin{equation}
\sigma^{*}=Ze\int_{x}^{\infty}N(x')dx'=\sigma\left[1+\frac{qx}{(2-q)\lambda}\right]^{1-\frac{2}{q}} \label{app}
\end{equation}

It is worthy noting that for strong coupling with $q/2 \ll 1$ and close to the macroion surface 
$x/\lambda \ll 2/q$ the counterion concentration profile,
Eq. (\ref{profile}) can be approximated by
\begin{equation}
N(x) \approx \frac{1}{2\pi l\lambda^{2}} \exp\left(-\frac{2x}{(2-q)\lambda}\right) 
\approx \frac{1}{2\pi l\lambda^{2}} \exp\left[-\frac{x}{\lambda}\left(1+\frac{q}{2}-...\right)\right]
\label{profilexs}
\end{equation}
We see that the strong coupling solution is the leading term of this expansion. 
This behavior at low distance from the surface is totally consistent with previous heuristic considerations \cite{EJPD}.
At large distances from the macroion surface, 
$x=x_{0}+\widetilde{x}$ (being $\widetilde{x} \ll x_{0}$, $x_{0}$ large) we have approximately
\begin{equation}
N(x) \approx N(x_{0})\left(\frac{x}{x_{0}}\right)^{-2/q} \approx N(x_{0})\left(1+\frac{\widetilde{x}}{\Lambda}\right)^{-2}
\end{equation}
with $\Lambda = x_{0}q$. PB behavior is 
thus regained if the surface is taken as if it were at $x_{0}$. When lowering $q$, 
the position $x_{0}$ must increase in order to be valid the expansion outlined. 
In the limit $q \to 0$, $x_{0}$ goes to infinity. 
This asymptotic behavior is consistent with PS theory \cite{Perel, EJPD} in which 
the boundary condition is displaced. There is no previous analytical theory, however, capable of reproducing 
the crossover between strong coupling and PB like behaviors \cite{Moreira2, EJPD}.
We achieve this here, through Eq. (\ref{profile}), for arbitrary coupling strength and position.

\section{Comparison with Monte Carlo simulations}

Moreira and Netz \cite{Moreira2} have performed extensive Monte Carlo calculations on the system under
consideration. These are based on the specification of the coupling constant $\Xi$, which enters in the 
Hamiltonian of their BG canonical partition function. The value of this constant in terms of the system
parameters is \cite{Moreira2}
\begin{equation}
\Xi=2\pi Z^{3} l_{B}^{2}\sigma \label{coupling}
\end{equation}
This coupling constant gives the strength of the coupling and depends on the surface charge density
of the macroion and on the ionic number $Z$ of the counterions.

\begin{figure}
\begin{center}
\includegraphics[angle=0, width=0.7\textwidth]{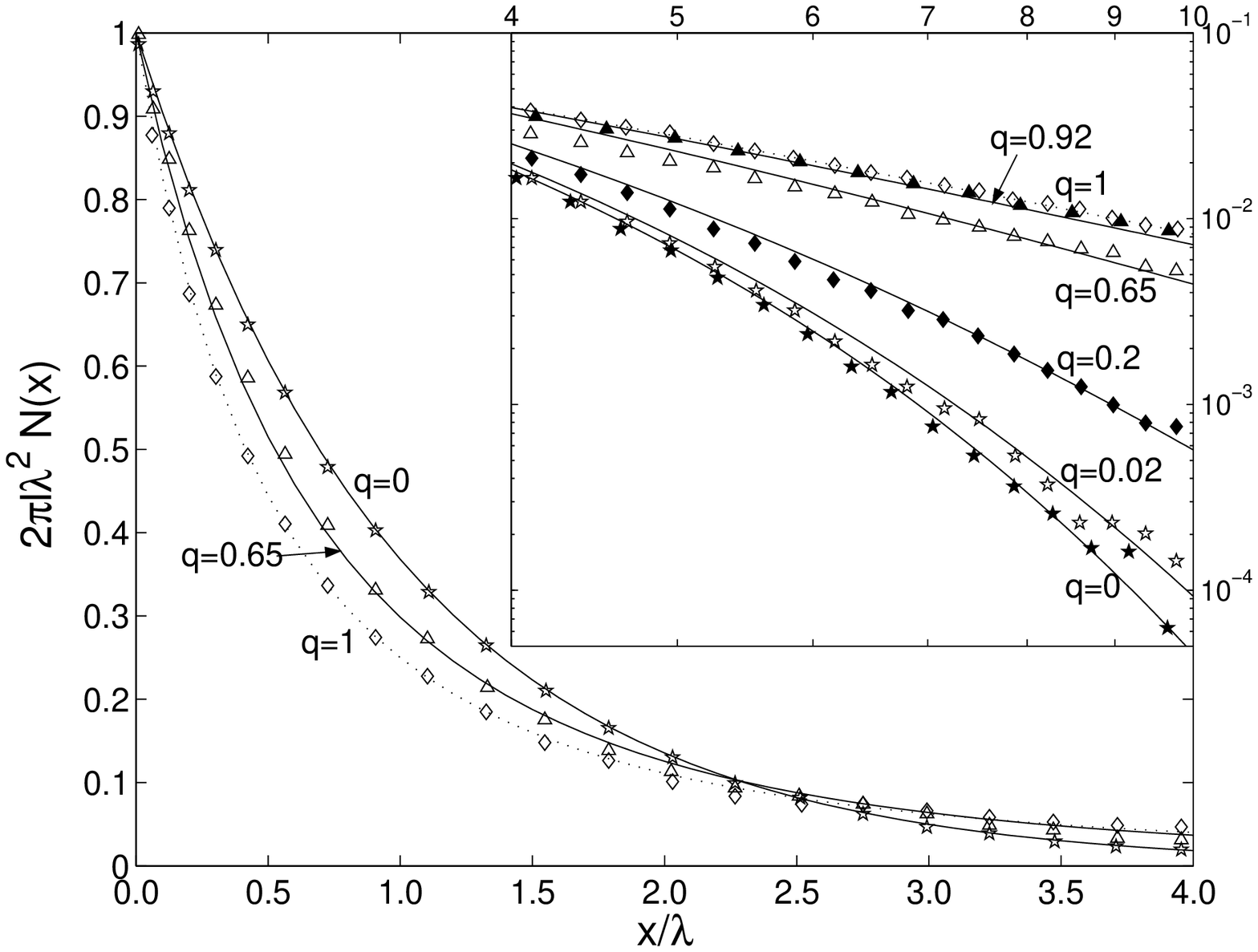}
\caption{}
\label{fig1}
\end{center}
\end{figure}

In Fig.\ref{fig1} we plot the counterion concentration profiles for the values of $q$ 
indicated in the figure.  Good fittings are obtained to Monte Carlo 
simulations by the authors of Ref. \cite{Moreira2} for different coupling strengths
and also for moderate coupling ($q=0.2$) where there exists no previous systematic theory 
to the best of our knowledge. The entropic parameter is directly related to the 
coupling constant considered in the simulations. In Table \ref{t1} we indicate
the values for the parameter $q$ related to the values of $\Xi$ in the simulations.

\begin{figure}
\begin{center}
\includegraphics[angle=0, width=0.7\textwidth]{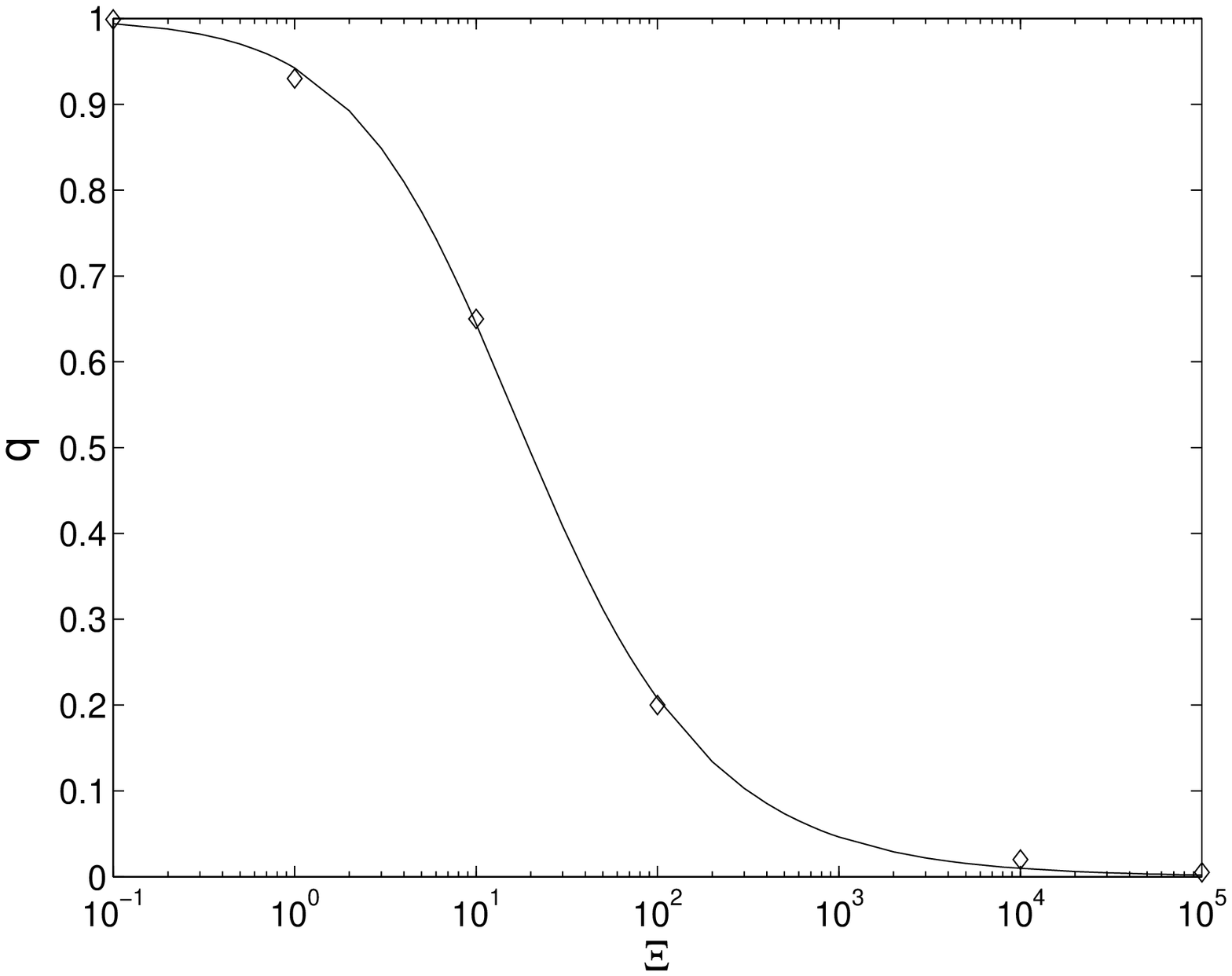}
\caption{}
\label{fig2}
\end{center}
\end{figure}

The data in Table \ref{t1} can be fitted accurately to a function $q=1/(1+a\Xi)^{b}$. We obtain
$a=0.091 \pm 0.015$ and $b=0.68 \pm 0.07$. These results are plotted in Fig. \ref{fig2}.
This provides a simple empirical expression relating the coupling constant $\Xi$ 
and the Tsallis entropic parameter $q$. The limit $q \to 0$ corresponds to a strongly correlated
liquid which is able to achieve a high degree of order at the vicinity of the macroion surface.
The weak coupling limit, reproduced taking $q \to 1$ corresponds to a disordered liquid lacking of
this ability to achieve order. It is worthy noting that the effect of the increasing correlations
leads to structures in which certain microstates are privileged. These structures can be rigorously 
described by BG statistics, although this is a challenging mathematical problem in most cases
requiring powerful computational resources. 
Some of the methods existing in the literature \cite{Moreira2, Stevens, Marcelja} work accurately
in the strong or weak coupling regimes, which correspond, approximately, to each of the two plateaus
in Fig. \ref{fig2}. In this paper we have provided an alternative and unified method, which provides complementary
insight, providing accurate expressions for the counterion concentration profiles valid for 
arbitrary coupling. The tricky intermediate region is also
accurately accounted for in the theory presented. Coupling is economically introduced through the parameter $q$
which is related to the relevant constant controlling the strength of the pair interaction term in the Hamiltonian
\cite{Moreira2}. 

It is worthy insisting on the fact that the exact many particle BG canonical partition function 
is the appropriate tool for the problem addressed here. 
Previous works and methods cited above are quite respectable and do constitute rigorous approaches 
that have provided many relevant insights in the physics involved, although restricted to certain coupling regimes. 
We cast no doubt on the validity and suitability of these methods. The issue addressed here is a complimentary
view that provides additional insight, approximating the exact many-particle BG distributions mean field
single-particle Tsallis distributions. Concretely, closed analytical expressions for the counterion 
concentration profiles valid for the whole spatial range and arbitrary coupling have been provided in excellent
agreement with previous numerical results. Previously calculated analytical limits are also exactly reproduced 
by the theory.

In the preceding treatment we have considered pointlike counterions. 
The theory can be applied to other geometries through the correct choice of the Laplacian in Eq. (\ref{GPBE})
and to a wide range of problems as, for example, polyelectrolyte multilayers. 
These questions, as well as studying charge inversion in the present framework, 
will be the aim of future work. 

We want to acknowledge an anonymous referee for helpful comments.
Financial support from the MCYT (Ministry of Science and Technology of Spain) and FEDER
under Project No.MAT2002-00646 is also gratefully acknowledged.

\begin{table}
\begin{center}
\begin{tabular}{c c}
\hline
$\Xi$ & $q$ \\
\hline
0.1  & 0.99\\
1  & 0.92\\
10  & 0.65\\
100  & 0.2\\
$10^{4}$ & 0.02\\
$10^{5}$ & 0.005\\
\hline
\end{tabular}
\caption{Values of $q$ corresponding to those of the coupling constant $\Xi$
considered in Ref. \cite{Moreira2}.}
\label{t1}
\end{center}
\end{table}

\newpage
\section*{Figure Captions}

\begin{enumerate}
\item{Counterion concentration profiles in dimensionless variables. 
Symbols correspond to the results of Monte Carlo simulations from Ref.\cite{Moreira2}.
The lines have been calculated from Eq.(\ref{profile}) 
for the values of $q$ indicated in the figure.
The dotted line corresponds to the solution of the PB equation ($q=1$).
The inset shows, in logarithmic scale, curves for increasing value of the dimensionless position
where other cases of moderate coupling have been added.}

\item{Comparison between data taken from Table \ref{t1} and a function $q=1/(1+a\Xi)^{b}$
for $a=0.091$ and $b=0.68$.}

\end{enumerate}

\end{document}